\documentclass[reprint,aps,pra,showpacs,superscriptaddress]{revtex4-1}
\usepackage{amssymb,amsmath}
\usepackage{graphicx}
\usepackage{dcolumn}
\usepackage{multirow}
\usepackage{color}
\usepackage[english]{babel}
\usepackage{hyperref}
\usepackage{cases}

\begin{document}

\def\simlt{\mathrel{\lower .3ex \rlap{$\sim$}\raise .5ex \hbox{$<$}}}
\def\simgt{\mathrel{\lower .3ex \rlap{$\sim$}\raise .5ex \hbox{$>$}}}

\title{\textbf{\fontfamily{phv}\selectfont Power law scaling for the adiabatic algorithm for search engine ranking}}
\author{Adam Frees}
\email{adam\_frees@brown.edu}
\affiliation{Department of Physics, Brown University, Providence, RI 02912}
\author{John King Gamble}
\email{jgamble@wisc.edu}
\affiliation{Department of Physics, University of Wisconsin-Madison, Madison, WI 53706}
\author{Kenneth Rudinger}
\affiliation{Department of Physics, University of Wisconsin-Madison, Madison, WI 53706}
\author{Eric Bach}
\affiliation{Department of Computer Sciences, University of Wisconsin-Madison, Madison, WI 53706}
\author{Mark Friesen}
\affiliation{Department of Physics, University of Wisconsin-Madison, Madison, WI 53706}
\author{Robert Joynt}
\affiliation{Department of Physics, University of Wisconsin-Madison, Madison, WI 53706}
\author{S. N. Coppersmith}
\email{snc@physics.wisc.edu}
\affiliation{Department of Physics, University of Wisconsin-Madison, Madison, WI 53706}


\begin{abstract}
An important method for search engine result ranking works by finding the principal eigenvector of the ``Google matrix." 
Recently, a quantum algorithm for preparing this eigenvector and evidence of an exponential speedup for some scale-free networks were presented.
Here, we show that the run-time depends on features of the graphs other than the degree distribution, and can be altered sufficiently to rule out a general exponential speedup.
For a sample of graphs with degree distributions that more closely resemble the Web than in previous work, the proposed algorithm for eigenvector preparation does not appear to run exponentially faster than the classical case.
\end{abstract}

\maketitle

\emph{Introduction.}---Quantum algorithms, which run on quantum computers, are known to be able to outperform classical algorithms for certain computational problems \cite{Shor1994, Grover1996}. Thus, finding a new algorithm that exhibits a quantum speedup, in particular an exponential speedup, is of great interest \cite{Bacon:2010:RPQ:1646353.1646375}. An extremely important problem in computer science is calculating ranking for search engine results. PageRank, first proposed by Brin and Page \cite{Brin:1998p107} underlies the success of the Google search engine \cite{Berhkin2005}. In this algorithm, websites are represented as nodes on a network graph, connected by directed edges that represent links. 
The matrix of network connections is constructed, and the PageRank vector is its principal eigenvector. Currently, computing the PageRank vector requires a time $O(n)$, where $n$ is the number of websites in the network considered (e.g. the World Wide Web) \cite{Garnerone:2012p230506}. Obtaining a quantum algorithm for PageRank that runs exponentially faster than the classical algorithm would be of great interest.

Recently, Garnerone, Zanardi, and Lidar (GZL) proposed an adiabatic quantum algorithm \cite{Farhi:2000preprint} to prepare the PageRank vector for a given network \cite{Garnerone:2012p230506}.
 Remarkably, GZL present evidence that this algorithm can prepare the PageRank vector in time $O \left[ \mathrm{polylog}(n)\right]$, exponentially faster than classical algorithms for certain networks.
 This runtime is due to the apparent logarithmic scaling of the gap between the two smallest eigenvalues of the Hamiltonian used in the algorithm (the energy gap). This scaling emerged on graphs constructed using adapted versions of two established methods of network construction: the preferential attachment model \cite{Barabasi:199p509} and the copying model \cite{Bollobas:2001p279}. Both of these models yield graphs that are similar to the connectivity of the World Wide Web in that they are sparse (the total number of edges scales at most proportionally to the number of nodes) and scale-free (the probability of finding a node with a specified in- or out-degree scales as a power law in those degrees). These features lead to networks that exhibit large-scale structure similar to that of the internet, such as being small-world \cite{Cohen:2003p058701} and loosely hierarchical \cite{Tangmunarunkit:2002p147}. 
GZL studied sets of networks that exhibited both logarithmic scaling and polynomial scaling of the gap in the system size.
However, they did not demonstrate that the networks with the favorable logarithmic gap scaling are scale-free over the region studied numerically.
 

Here, we study the scaling of the GZL algorithm for graphs with degree distributions consistent with the internet.
A realistic network model of the World Wide Web must be scale-free in both the in- and the out-degree \cite{Bollobas:2003p132,Broder:2000p309}. 
We consider a broad variety of scale-free networks constructed by different methods.
Choosing three well-known models for constructing random, scale-free networks, we control for both the mean degree and the exponent of the power-law governing the degree distribution.
We find that graphs with the same degree distribution can have different energy gap and run-time behaviors. 
Finally, we focus on degree distributions described by power laws consistent with those measured for the Web, both for the in-degree and the out-degree. We find that the relevant energy gap scales as a power of the system size, rather than logarithmically.
These results demonstrate that for Web-like graphs, the GZL adiabatic algorithm does not yield an exponential quantum speedup for preparing the PageRank vector compared to current classical algorithms.



\begin{figure}[tb]
\includegraphics[width=1.0 \linewidth]{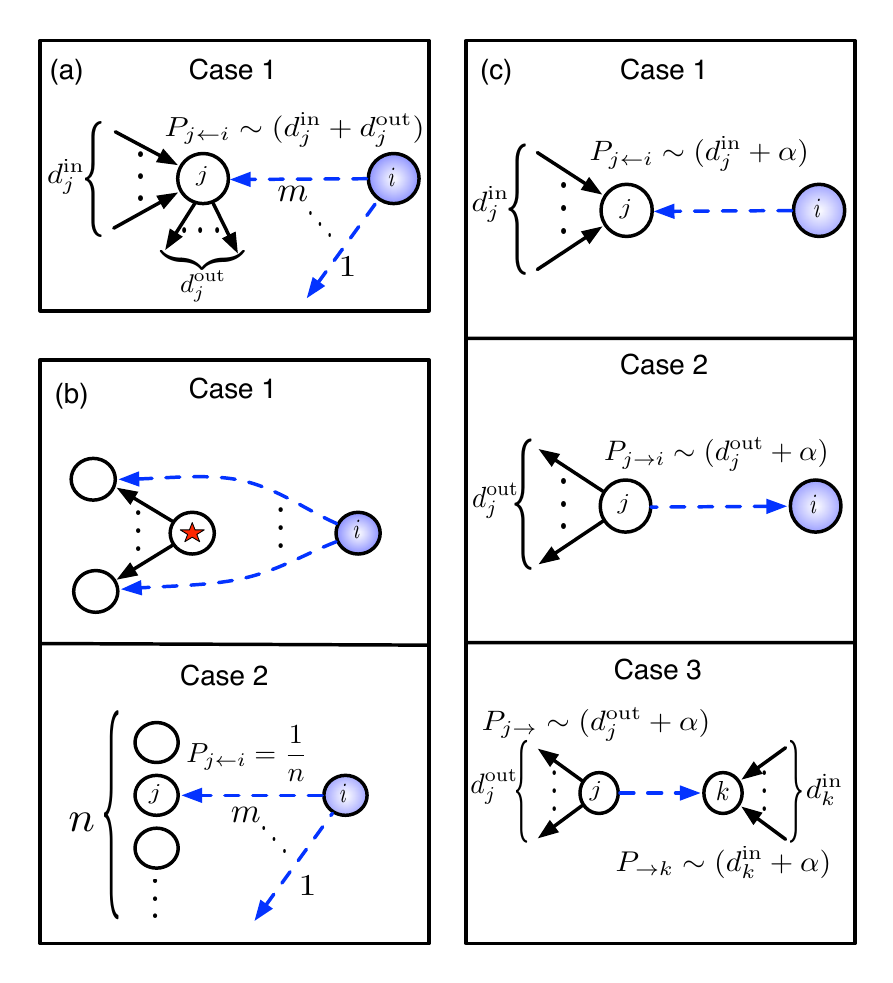} 
\caption{(color online) Illustrations of the three network generation models used. (a): GZL  \cite{Garnerone:2012p230506} preferential attachment, (b): GZL copying, and (c): $\alpha$-preferential attachment \cite{Bollobas:2003p132,Chung:2006}. In all three models, a network is constructed by adding vertices and edges sequentially. (a): At each time step a new vertex $i$ is added with $m$ outgoing edges. The probability that one of these edges connects to a node $j$ is proportional to the total degree of $j$. (b): At each time step there are two possible actions. With probability $(1-p)$, the new vertex points to all of the same vertices as the ``star vertex," which is a pre-existing vertex chosen uniformly at random at each time step. With probability $p$, $m$ outgoing edges are added to the new vertex, each pointing to vertices chosen uniformly at random. (c): There are three possible actions at each time step. With probability $p_1$, a new vertex is added with a single outgoing edge, pointing to a node $j$ with probability proportional to the in-degree of $j$ plus a parameter $\alpha$. With probability $p_2$, a new vertex is added with a single incoming edge, pointing from a node $j$ with probability proportional to the out-degree of $j$ plus $\alpha$. With probability $(1-p_1-p_2)$, no vertex, only an edge, is added. Its ending and starting points are determined as in cases 1 and 2, respectively. In all panels, the newly-added edges are indicated by dashed lines.\label{fig1}}
\end{figure}

\emph{Network growth models.}---We generate samples of graphs with prescribed degree distributions using three different network growth models.
GZL \cite{Garnerone:2012p230506} use modified versions of two network construction algorithms: the preferential attachment model \cite{Barabasi:199p509}  and the copying model \cite{Bollobas:2001p279}. In addition to these two models, here we include also the more complex $\alpha$-preferential attachment model described by Bollob\'{a}s \emph{et al.} \cite{Bollobas:2003p132,Chung:2006}. All three models grow random networks using probabilistic rules at discrete construction steps, which are detailed in Fig.~\ref{fig1}. 

All three of these models produce sparse, scale-free directed networks, in which  the probability of the in-degree (the number of incoming edges) and out-degree (the number of outgoing edges) of node $i$ being equal to $k$ are each proportional to a power law: 
\begin{align}
P(d_{in}(i) &= k) \sim k^{-\gamma_{in}} \label{Gamma_In}\\
P(d_{out}(i) &= k) \sim k^{-\gamma_{out}} \label{Gamma_Out},
\end{align}
where $d_{in}(i)$ and $d_{out}(i)$ are the in- and out-degrees of node $i$, respectively, and the exponents $\gamma_{in}$ and $\gamma_{out}$ are typically between $2$ and $4$ \cite{Barabasi:199p509}.
The GZL versions \cite{Garnerone:2012p230506} of the preferential attachment and copying models \cite{Barabasi:199p509, Kleinberg:1999p1} produce networks that are scale-free in the limit of large graph size. However, due to the addition procedure described below, the networks are not necessarily scale-free for the sizes of graphs studied numerically here and in Ref. \cite{Garnerone:2012p230506}.
To achieve networks that are scale-free in the out-degree, GZL suggest to
construct two networks, $X$ and $Y$, independently.  $X$ and $Y$ are each generated as in Fig.~\ref{fig1}, except that for $Y$ the direction of the edges added is reversed.
The networks can then be added together, and the weights and loops discarded \cite{Garnerone:2012p230506,Silvano:PrivateComm2012}.  The resulting composite network is scale-free in both in-degree and out-degree, provided $X$ and $Y$ have the same number of edges per node.  (See Supplemental Materials for details \cite{supp_info}.)
In contrast to Ref. \cite{Garnerone:2012p230506}, the graphs studied here are all constrained in this way.  However, the graphs exhibiting logarithmic scaling in \cite{Garnerone:2012p230506} are not \cite{Silvano:PrivateComm2012}, and they do not exhibit truly scale-free degree distributions over the numerically studied region.
On the other hand, the $\alpha$-preferential attachment model (considered here but not in \cite{Garnerone:2012p230506}) constructs a network which is scale-free in both in- and out-degrees without requiring an additional combination step.  As with the GZL preferential attachment model, all weights and loops are removed from the final $\alpha$-preferential attachment network.

The exponents $\gamma_{in}$ (Eq.~\ref{Gamma_In}) and $\gamma_{out}$ (Eq.~\ref{Gamma_Out}) of the degree distribution are model-dependent. In the GZL preferential attachment model the number of edges added at each construction step controls the sparsity, and it is always the case that $\gamma_{in} = \gamma_{out} = 3$ \cite{Barabasi:199p509}.
 Both the GZL copying model and $\alpha$-preferential attachment allow for independently tunable exponents and mean degree.  (See Appendix \ref{web_graph_parameters} for details.) This flexibility enables us to create three ensembles of model networks that have nearly identical degree distributions for $\gamma_{in} = \gamma_{out} = 3$. Further, the last two models can be set with the exponents estimated for the World Wide Web \cite{Bollobas:2001p279, Bollobas:2003p132}, namely $\gamma_{in} = 2.1$ and $\gamma_{out} = 2.72$ \cite{Broder:2000p309}.

\emph{Algorithm description.}---The Google matrix is constructed by taking as input an unweighted, simple network with $n$ nodes \cite{Brin:1998p107}, and representing it as an adjacency matrix $A$, where $A(i,j)=1$ if a directed edge points from node $i$ to node $j$, and $0$ otherwise.  From this, one defines the matrix $P$:
\begin{subnumcases}{P(i,j) = }
1/d_{out}(i) & if $A(i,j) = 1$\label{P_def_1}\\
1/n & if $\forall j, A(i,j) = 0$\label{P_def_2}\\
0 & otherwise\label{P_def_3}
\end{subnumcases}
The matrix $P$ is stochastic because $\sum_j P(i,j) = 1$ for all $i$. $P$ can be thought of as a random walk (i.e. a web-surfer), where the walker follows the network with equal likelihood of traversing all allowed links. If the walker ever reaches a dangling node (a node with $d_{out} = 0$), Eq.~\ref{P_def_2} implies that it can randomly hop to any vertex with equal probability. To prevent the walker from becoming trapped in an isolated portion of the network (a sink), the probability $(1-\alpha_g)$ of moving to a node uniformly at random (including the possibility of staying still) is included, where $0<\alpha_g<1$; Google uses $\alpha_g = 0.85$, which we also use here \cite{Garnerone:2012p230506}. The Google matrix $G$ is defined as the transpose of this resulting transition matrix:

\begin{equation}
G = \alpha_g P^T + (1 - \alpha_g) J,
\end{equation}
where $J$ is the matrix of all ones. The PageRank vector $\vec p$ is the unique eigenvector associated with the largest eigenvalue of $G$, which is $1$. The runtime of the best classical algorithm, which calculates the PageRank vector via power iteration, is $O(n)$ \cite{Brin:1998p107, Garnerone:2012p230506}.

To formulate an adiabatic quantum algorithm, GZL construct the Hamiltonian $h(G)$:
\begin{equation}
h(G) = \left(\mathbb I - G \right)^\dagger \left(\mathbb I - G \right),
\end{equation}
which is Hermitian, even though $G$ is not. The ground state of this Hamiltonian is the normalized PageRank vector. The adiabatic algorithm is completely defined by the interpolation Hamiltonian $H(s) =  s  h(G)+ (1-s) h(G_c) $, where $s\in [0,1]$, and $G_c$ is the Google matrix for the complete network (including loops), whose ground state is a uniform superposition. The adiabatic theorem guarantees that if we initialize our system in the ground state of $h(G_c)$ and change $s$ from $0$ to $1$ sufficiently slowly, the system remains in the ground state \cite{Farhi:2000preprint}.  Since the PageRank vector is the ground state of $H(1) = h(G)$, the PageRank vector is obtained when $s=1$. The required slowness is also determined by the adiabatic theorem: as long as $s(t)$ is a smooth function of the time $t$ with $0\leq t \leq T$, the runtime $T \sim \delta^{-b}$, where $b$ is $O(1)$ and $\delta$ is the energy gap between the ground and first excited state of $H(s)$, minimized over $s$ \cite{Farhi:2000preprint}. Thus, an exponential speedup over the classical case is possible if $\delta^{-1}$ is $O[\log(n)]$, since then $T$ is $O \left[ \mathrm{polylog}(n)\right]$.

\begin{figure}[tb!]
\includegraphics[width=1.0 \linewidth]{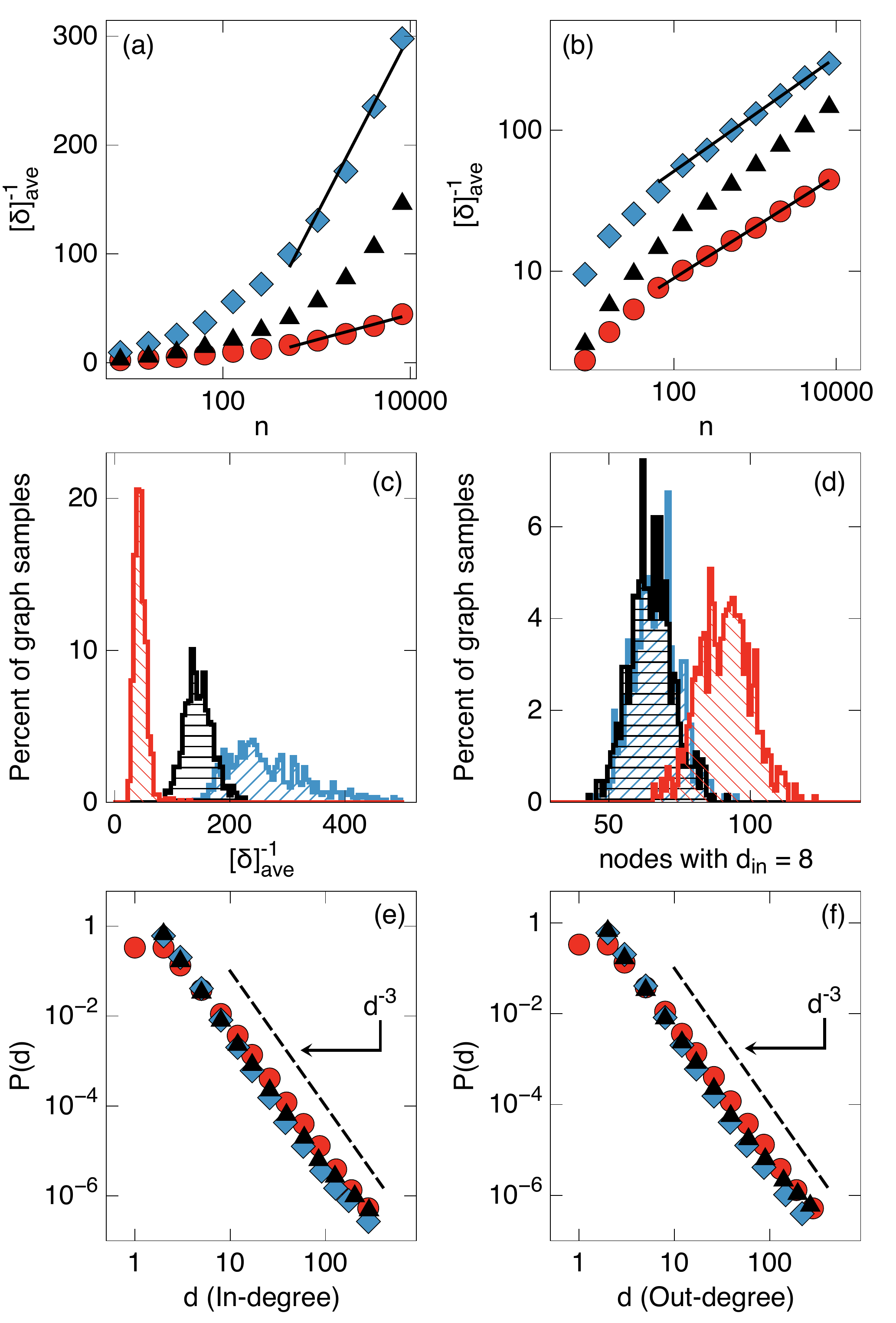} 
\caption{(color online) Comparison of the scaling of the inverse energy gap $\delta^{-1}$ for the GZL \cite{Garnerone:2012p230506} preferential attachment model (triangles, horizontal hatching), GZL copying model (diamonds, upward-sloping hatching), and $\alpha$-preferential attachment model  \cite{Bollobas:2003p132} (circles, downward-sloping hatching), shown on (a): Semilog and (b): Log-Log scales, demonstrating that $\delta^{-1}$ is not proportional to $\log{(n)}$ for these models. Results are averaged over 1000 random instances for $n<8192$, and over 500 random instances at $n=8192$. The fitting lines showed in (a) are $72.2\cdot \ln(n)- 363$ for the copying model and $10.1\cdot\ln(n) - 48.8$ for the $\alpha$-preferential attachment model. 
In (b), the fits shown are $8.0 \cdot n^{0.4}$ for the copying model and $1.7 \cdot n^{0.4}$ for the $\alpha$-preferential attachment model. 
If we fit the data instead to a power of a logarithm (not shown), we obtain $ 0.56\cdot \ln^{2.9}(n)$ for the copying model and $0.18 \cdot \ln^{2.5}(n)$ for the  $\alpha$-preferential attachment model. (c): Histogram of the inverse energy gaps for the data shown in panels (a)-(b) at $n=8192$. (d): Histogram showing the distribution of number of vertices with in-degree $d_{in} = 8$ for $n=8192$. (e)-(f): Degree-distributions of the three models, demonstrating scale-free behavior and indicating that $\gamma_{\text{in}} = \gamma_{\text{out}} = 3$. Adaptive binning was used, as described in Appendix \ref{adaptive_binning}. In all cases, both the mean in- and out-degree of each graph are $2$ edges per node. These results demonstrate that $\delta^{-1}$ differs significantly for the different graph construction methods, while the degree distributions are very similar.}\label{fig2} 
\end{figure}

\begin{figure}[tb]
\includegraphics[width=1.0 \linewidth]{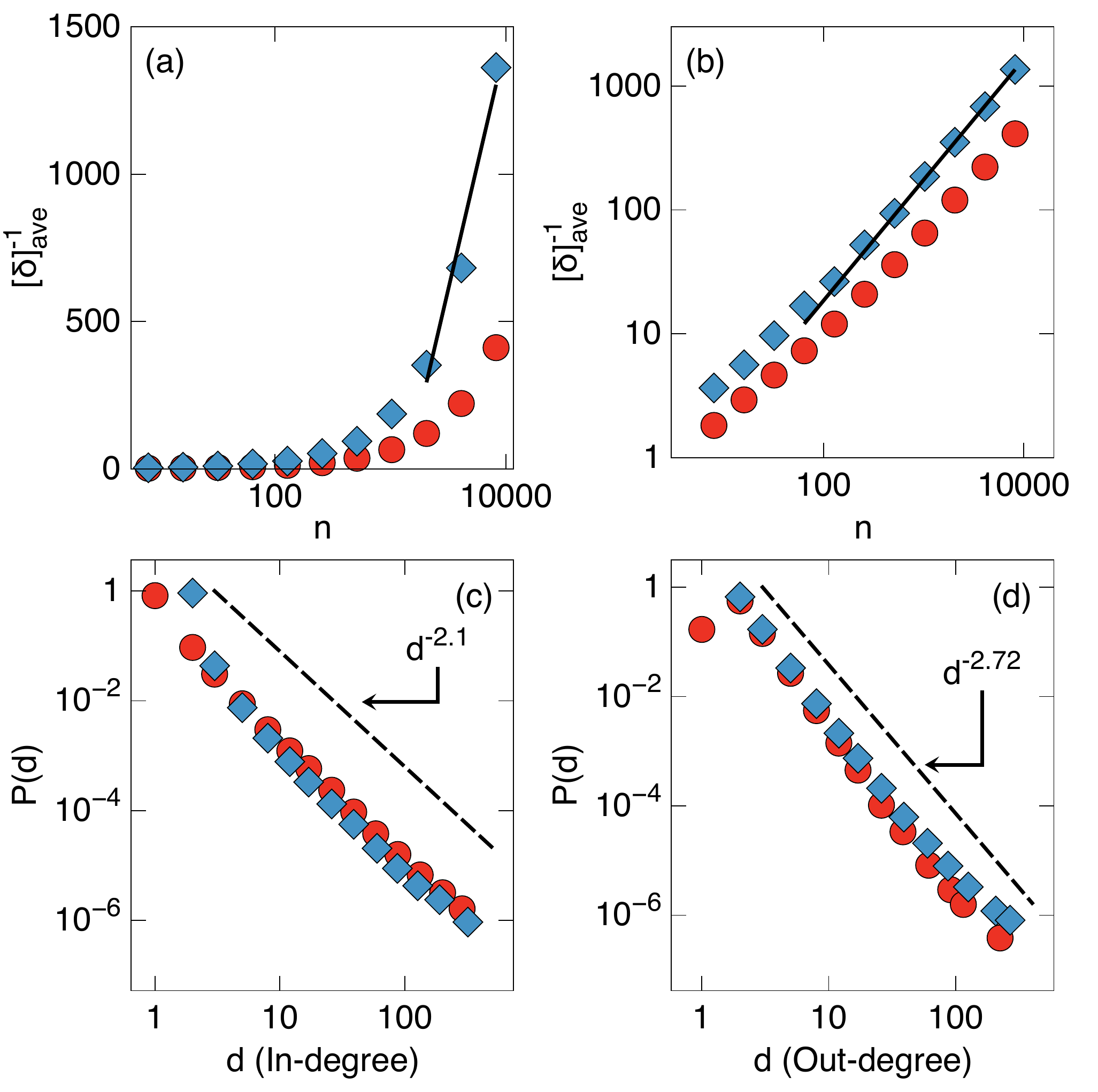} 
\caption{(color online) Inverse energy gap scaling for GZL \cite{Garnerone:2012p230506} copying model (diamonds), and $\alpha$-preferential attachment model  \cite{Bollobas:2003p132} (circles) of WWW-like networks, shown on (a): Semilog and (b): Log-Log scales. Results are averaged over 1000 random instances for $n<8192$, and over 500 random instances at $n=8192$. In (a), the line fit shown is $730 \cdot \ln(n) - 5300$, while in (b) the line fit is $0.2\cdot n^{0.97}$.
If we fit the data to a power of a logarithm (not shown), for the copying model we obtain $ 3 \times 10^{-5} \cdot \ln^{8.0}(n)$. Because of the large power of the logarithm required for the polylogarithmic fit, the power-law dependence on $n$ appears more natural and plausible.
(c)-(d): Degree-distributions of the two models, histogrammed using adaptive binning (see Appendix \ref{adaptive_binning}), indicating that $\gamma_{\text{in}} = 2.1$ and $\gamma_{\text{out}} = 2.72$, corresponding to the estimates for the degree distribution of the World Wide Web \cite{Broder:2000p309}. In all cases, the mean in- and out-degree of each network were each $2$ edges per node.
\label{fig3}}
\end{figure}

\emph{Numerical results.}---To study the scaling of the minimum energy gap $\delta$ with the network size $n$, we compute $\delta$ for the GZL Hamiltonian $H(s)$, averaging the results over many network realizations (typically 1000). Specifically, we calculate the minimum value of $\delta$ over $s \in [0,1]$ using the Nelder-Mead method \cite{Nelder1965}, where each objective function call calculates directly the eigenvalue spectrum of $H(s)$. We find that for most, but not all, network choices the minimum gap occurs when $s=1$. Since $H(s)$ is a dense matrix, this process is computationally intensive.
We use the University of Wisconsin-Madison Center for High Throughput Computing and Open Science Grid to perform the simulations.

To assess whether the inverse energy gap $\delta^{-1}$ scales logarithmically or as a power-law in $n$, we plot in Fig.~\ref{fig2} $\delta^{-1}$ versus the network size on both log-linear and log-log scales, with data for the GZL preferential attachment, GZL copying, and $\alpha$-preferential attachment models. The model parameters  are tuned (see Appendix \ref{web_graph_parameters}) so that all three have $\gamma_{in}=\gamma_{out}=3$ and have an average of $2$ in- and $2$ out-edges per node. Despite having nearly identical degree distributions (shown in Figs.~\ref{fig2}(e) and \ref{fig2}(f)), the scaling of $\delta^{-1}$ depends significantly on the method used to construct the graphs when viewed in Fig.~\ref{fig2}(a). 
In Fig.~\ref{fig2}(c), we show the distribution corresponding to the final data points in  Fig.~\ref{fig2}(a), where we see that the distributions are well-separated and hence the construction models give different values of $\delta^{-1}$.
By contrast, the degree distributions are difficult to distinguish, as shown in Fig.~\ref{fig2}(d).
When viewed in Fig.~\ref{fig2}(b), the scaling of $\delta^{-1}$ is similar for all three methods of graph construction. The data in Fig.~\ref{fig2} clearly do not scale linearly with $\log{(n)}$. We conclude that the data are more consistent with $\delta^{-1}$ scaling either polylogarithmically or as a power law, rather than logarithmically. 

We next perform a similar analysis for degree distributions more closely related to the network of primary interest, the World Wide Web, for which a realistic set of degree parameters is given by $\gamma_{in} = 2.1$ and $\gamma_{out} = 2.72$ \cite{Broder:2000p309}. As mentioned above, the preferential attachment model cannot be tuned to obtain degree parameters other than $3$. However, the other two network models can be adjusted to match these values \cite{Bollobas:2001p279, Bollobas:2003p132}. 
More details on this are discussed in Appendix \ref{web_graph_parameters}.
As before, we set the mean degree to be $2$ in- and $2$ out-edges per node. 

Fig.~\ref{fig3} presents the results of these simulations, clearly indicating that $\delta^{-1}$ scales at least as a power of $n$. In particular, we note that the prefactor of the logarithmic fit is over $700$ and the power of the logarithm in the polylogarithmic fit is 8, while the power law fit exponent is close to one. The results do not change substantially when the mean degree is varied and the degree distributions exponents are fixed. These data indicate that for graphs with degree distributions similar to those measured for the World Wide Web, the GZL adiabatic algorithm for PageRank vector preparation is unlikely to provide an exponential speedup over the classical case.

\emph{Discussion.}---We have investigated the recently proposed adiabatic quantum algorithm for preparing the PageRank vector using an adiabatic quantum algorithm \cite{Garnerone:2012p230506}. We find that the eigenvalue gap that determines the algorithm runtime depends on the method of construction of the network, even when the feature believed to be critical for large-scale network structure, the degree distribution, is held fixed. The exponent governing the variation of the gap with graph size does not vary significantly with the method of construction only if power-law scaling of the gap with size is assumed. 
For networks that are scale-free in their in- and out-degree distributions, and particularly when the degree distributions similar to those measured for the World Wide Web, our numerical results indicate strongly that the GZL adiabatic algorithm for PageRank vector preparation does not offer an exponential speedup over current classical algorithms.


This work was supported in part by ARO, DOD (W911NF-09-1-0439) and NSF (CCR-0635355, DMR 0906951). AF acknowledges support from the NSF REU program (PHY-PIF-1104660). We thank S. Garnerone, D. A. Lidar, and D. Bradley for useful discussions.  We also thank the HEP, Condor, and CHTC groups at University of Wisconsin-Madison for computational support.

\appendix

\section{Parameters of Web Graph Models}\label{web_graph_parameters}
In implementing the models used in this paper, the relationship between the parameters of the network generation algorithms and the generated networks themselves is not always obvious, so in the following section we explain it in detail.
\subsection{GZL Preferential Attachment}
The method of graph construction in the GZL Preferential Attachment Model \cite{Garnerone:2012p230506} consists of two phases, each with its own parameter. First, a graph $X$ (with adjacency matrix $A_X$) is created by adding a new vertex at each time step, where each vertex is created with $m_{X}$ out-going edges. Next, a second graph $Y$ (with adjacency matrix $A_Y$) is created in the same fashion, only with each new vertex having $m_{Y}$ in-coming edges. $A_X$ and $A_Y$ are then added together, with loops and weights discarded, forming the adjacency matrix of the desired network. $m_{X}$ and $m_{Y}$ are the two parameters to consider in this algorithm.

In order for a graph to be scale-free, $\text{Pr}(d_{in}=k)$ and $\text{Pr}(d_{out}=k)$, the probabilities that the in-degree $d_{in}$ and the out-degree $d_{out}$ of a random node have the value $k$, must satisfy
\begin{align}\label{ideal_distro}
\text{Pr}(d_{in}&=k) \sim k^{-\gamma_{in}},\\
\text{Pr}(d_{out}&=k) \sim k^{-\gamma_{out}} , \nonumber
\end{align}
where $\gamma_{in}$ and $\gamma_{out}$ are positive real numbers, and it is understood that $\mathrm{Pr}(d_{in}=k) = 0$ when $k<m_X$ and $\mathrm{Pr}(d_{out}=k) = 0$ when $k<m_Y$. 
To compute $\gamma_{in}$ and $\gamma_{out}$, one starts from the undirected version from Ref. \cite{Barabasi:1999p173}.
This result is then combined with a constant offset, since each vertex of $X$ has $m_x$ outgoing edges and each vertex of $Y$ has $m_Y$ incoming edges. 
The resulting composite probability distributions follow
\begin{align}\label{actual_distro}
\text{Pr}(d_{in}&=k) \sim (k + m_X - m_Y)^{-3}, \\ 
\text{Pr}(d_{out}&=k) \sim (k - m_X + m_Y )^{-3}.  \nonumber
\end{align}
Thus, for sufficiently large $k$, these distributions are scale-free. However, for a large range of intermediate $k$, we expect substantial deviation from the power law dependence of Eq.~(\ref{ideal_distro}).
According to GZL \cite{Silvano:PrivateComm2012}, the parameters used to generate Fig.~2 in their paper \cite{Garnerone:2012p230506}, which provides the main evidence for logarithmic scaling of the gap, follow $m_Y \gg m_X$.
In Fig.~\ref{fig4}, we show the degree distributions for such a network, where we set $m_X=1$ and $m_Y=15$. 
There, we see that the degree distributions are well-described by Eq.~({\ref{actual_distro}), and that the addition process does indeed distort the degree distributions.
By requiring $m_{X} = m_{Y}$, as we have done in this paper (and GZL did for a portion of their supplemental material \cite{Garnerone:2012p230506}), $\gamma_{in} = \gamma_{out}= 3$ for all $k$, meaning that the in-degrees and out-degrees both follow the desired power law behavior. 

\begin{figure}[tb]
\includegraphics[width=1.0 \linewidth]{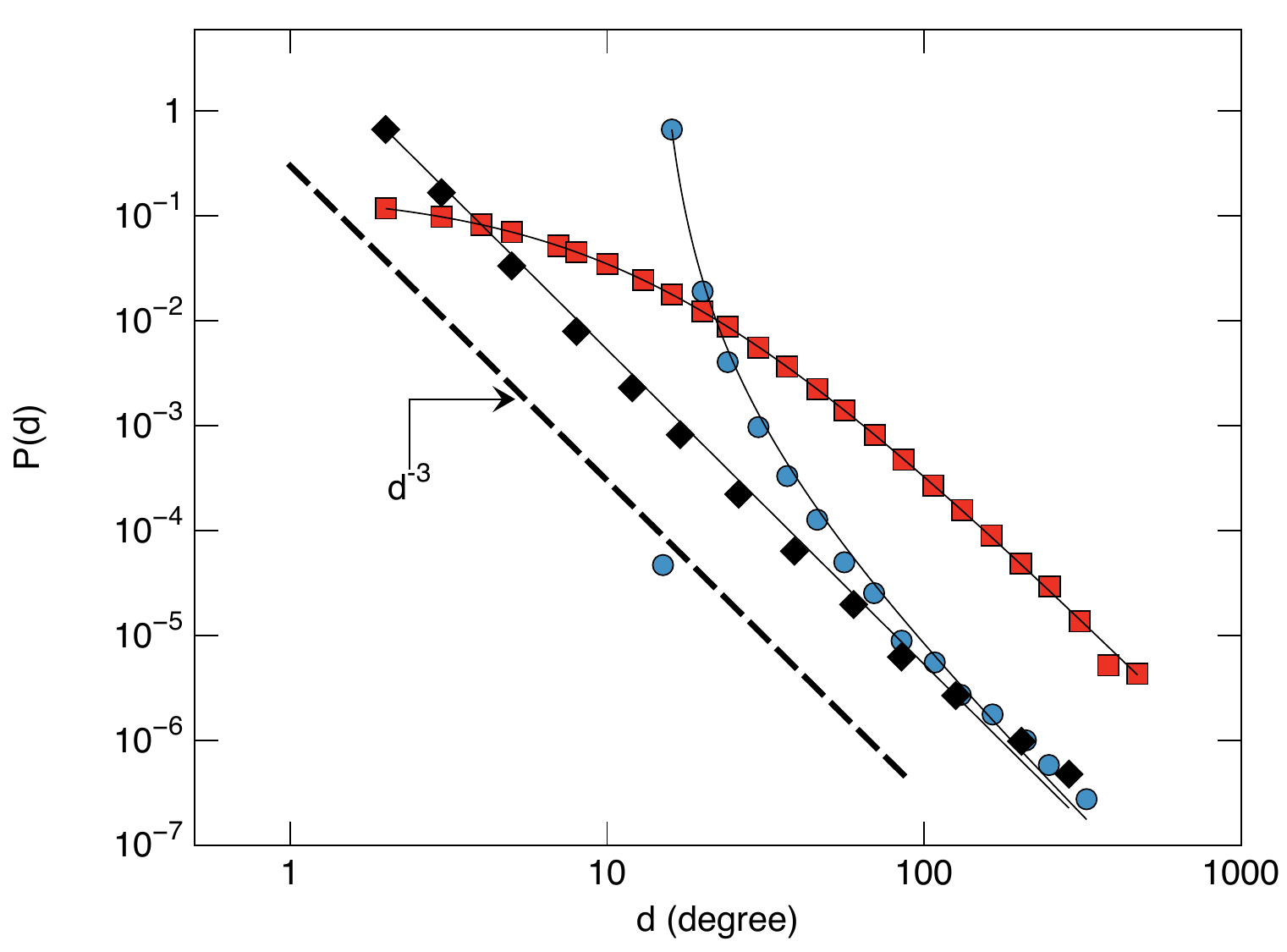} 
\caption{(color online) Degree distributions for the GZL preferential attachment model with $m_X = 1$ and $m_Y = 15$, taken at graph size $n=8196$ and averaged over approximately $1000$ random graph realizations. Both the in-degree (blue circles) and out-degree (red squares) distributions are shown. For reference, the in-degree distribution for $m_X = 1$ and $m_Y = 1$ (duplicated from Fig.~\ref{fig2} of the main text is shown (black diamonds). The dashed line is the expected power law scaling of $d^{-3}$, which is applicable for large $d$. As predicted by Eq.~(\ref{actual_distro}), shown as fitting curves, the $m_X=1$ and $m_Y=15$ distributions exhibit non-scale-free behavior over a wide region of $d$.
\label{fig4}}
\end{figure}

The asymptotic (large number of nodes) value of average edges per node for the composite graph is also determined by the parameters $m_{X}$ and $m_{Y}$. Because $m_X$ is the number of out-going edges per vertex in graph $X$, it is also the average number of edges per vertex in $X$. The same logic holds for $m_Y$ and graph $Y$. Thus, when constructing the composite graph, the asymptotic average edges per node would be simply $m_{X}+m_{Y}$. Although loops are then eliminated from the composite graph, the expected number of loops is much less than $n$ in the large-$n$ case, so this has little effect on the average edges per node. To produce a graph with $\gamma_{in} = \gamma_{out}= 3$ and average in- and out-edges per node of 2 (as in Fig. 2 of the main text), we use this model with $m_{X}=m_{Y}=1$.

\subsection{GZL Copying Model}
The parameters of the GZL Copying Model \cite{Garnerone:2012p230506} are similar to the GZL Preferential Attachment, as they both involve the adding of two graphs to form a composite graph. We again have the parameters $m_{X}$ and $m_{Y}$, which again indicate the number of out-going edges per node in one component graph and the number of in-coming edges per node in the other.

This model has two new parameters, $p_X$ and $p_Y$, which are the probabilities of a new node connecting to nodes chosen uniformly at random at a given time step during the construction of $X$ and $Y$, respectively. We follow Ref. \cite{Kleinberg:1999p1} and add a constant offset (just as in the preferential attachment case). Doing so, we again obtain the result that the graphs are scale-free only for $m_{X}=m_{Y}$.
Assuming this constraint, the composite graph follows
\begin{align}
\gamma_{in} = \frac{2-p_X }{1-p_X},\\
\gamma_{out} = \frac{2-p_Y }{1-p_Y}. 
\end{align}

For the data in Fig. 2 of the main text, we used the parameters $p_X = p_Y = 0.5$ and $m_X = m_Y = 1$. In Fig. 3 of the main text, we used $p_X = 1/11$ and $p_Y = 35/86$ and $m_X = m_Y = 1$.

\subsection{$\alpha$-Preferential Attachment}
Just as in the GZL Copying Model, there are multiple possible actions at each time step in the $\alpha$-Preferential Attachment Model \cite{Bollobas:2003p132}, and each of these steps has an associated probability. $p_1$ is the probability of adding a new vertex with a single out-going edge, $p_2$ is the probability of adding a new vertex with a single in-coming edge, and $1-p_1-p_2$ is the probability of an edge being added to the existing network without the addition of a new vertex. $\alpha$, the third parameter, measures how far the generated network deviates from the GZL preferential attachment model.

As laid out in Ref. \cite{Bollobas:2003p132}, the relationship between these 3 parameters and the exponents is

\begin{align}
\gamma_{in} = \frac{2+ (p_1 + p_2)\alpha-p_2 }{1-p_2},\\
\gamma_{out} = \frac{2 + (p_1 + p_2)\alpha-p_1 }{1-p_1}. 
\end{align}

The connection between these parameters and the average number of directed edges per node in the graph is clear when one considers that the probability that a new node will be added at a given time step is $p_1 + p_2$, and a new edge is added at each step.

Using these constraints, we can find appropriate values for the parameters for both Fig. 2 and Fig. 3 of the main text. In Fig. 2, we used $p_1 = p_2 = 0.25$, and $\alpha = 1$, and in Fig. 3, we used $p_1 = 0.415$, $p_2 = 0.0851$, and $\alpha = 0.0128$. These choices in parameters keep $\gamma_{in}$ and $\gamma_{out}$ fixed at our desired values, while simultaneously keeping the graph at an average of $2$ in- and $2$ out-edges per node.

\section{Initial Conditions}\label{initial_conditions}
For each of these models, it is necessary to specify an initial graph to seed the network growth. In our simulations we used a complete graph (including loops) with $m + 1$ vertices, where $m$ is the number of edges added per vertex (in the $\alpha$-Preferential Attachment Model, we used $m = 1$).

\section{Adaptive Binning}\label{adaptive_binning}
In the plots of the degree distributions (Figs.~2(e)-(f), Figs.~3(c)-(d), and Fig.~\ref{fig4}), numerical noise caused by few high-degree vertices leads to data which are difficult to interpret. In order to combat this, we use adaptive binning, which functions as follows. First, some sampling threshold $s_t$ is set, which we take to be $200$ in our analysis.
If a given data point, corresponding to a degree, contains at least $s_t$ samples, then it is included.
If the data point instead has fewer than $s_t$ samples, it is combined with nearby points until the aggregated samples total at least $s_t$. 
The weighted average degree and probability are then recorded.

\end{document}